\documentclass[checkin,showpacs,aps,prl]{revtex4}
\usepackage{threeparttable}
\usepackage{amsmath}
\usepackage{multirow}
\usepackage{booktabs}
\usepackage{graphicx,color}
\usepackage{graphicx}
\usepackage{subfigure}
\usepackage{amsmath}
\newcommand{\ba}{\begin{eqnarray}}
\newcommand{\ea}{\end{eqnarray}}

\newcommand{\ub}{\mu_{\rm B}}

\newcommand{\be}{\begin{equation}}
\newcommand{\ee}{\end{equation}}

\newcommand{\ra}{\rangle}
\newcommand{\et}{{\it et al. }}

\definecolor{pink}{rgb}{1,0.18,1.0} 
\def\prl{{ Phys. Rev. Lett. }}

\def\prb{{ Phys. Rev. B }}

\def\np{{Nature Phys. }}
\def\jpcm{{J. Phys.: Condens. Matter }}

\begin{document}

%\title{ Three-dimensional mapping of femtosecond magnetization in
%  crystal momentum space }

%\title{ Kernel of femtomagnetism in the crystal-momentum space:
%  \\ Manifestation of an optical spin generator }

\title{Hot spin spots in the laser-induced
    demagnetization }

\author{M. S. Si}
\affiliation{Department of Physics, Indiana State University, Terre
  Haute, Indiana 47809, \& \\ Key Laboratory for Magnetism and
  Magnetic Materials of the Ministry of Education, Lanzhou University,
  Lanzhou 730000, China}
 
\author{G. P. Zhang$^{*}$}
\affiliation{$^{1}$Department of Physics, Indiana State University, Terre
  Haute, Indiana 47809}

% However, surprisingly, nearly all the existing
%   investigations have only focused one contribution, which limits the
%   scope of the. 

\newcommand{\OSGe}{$\cal SD$}
\newcommand{\OSG}{$\cal SD$ }
\newcommand{\osg}{{\cal SD} }

\date{\today}

% Laser-induced femtosecond magnetism or femtomagnetism simultaneously
%   relies on two distinctive contributions to influence spin: One is
%   the interaction between a laser field and a magnetic system, and
%   the other is the intrinsic spin moment change among electronic
%   states. To understand femtomagnetism it is crucial to take into
%   account both contributions simultaneously.

\begin{abstract}
 { Laser-induced femtosecond magnetism or femtomagnetism
   simultaneously relies on two distinctive contributions: (a) the
   optical dipole interaction (ODI) between a laser
   field and a magnetic system and (b) the spin expectation value
   change (SEC) between two transition states.  Surprisingly, up to
   now, no study has taken both contributions into account
   simultaneously.  Here we do so by introducing a new concept of the
   optical spin generator, a product of SEC and ODI between transition
   states. In ferromagnetic nickel, our first-principles
     calculation demonstrates that the larger the value of optical
     spin generator is, the larger the dynamic spin moment change is.
     This simple generator directly links the time-dependent spin
     moment change $\Delta M_{z}^{\bf k}(t)$ at every crystal-momentum
     ${\bf k}$ point to its intrinsic electronic structure and
     magnetic properties.  Those hot spin spots are a
     direct manifestation of the optical spin generator, and should be
     the focus of future research.}  
\end{abstract}
 \maketitle
%\section{I. introduction}
 
Femtomagnetism represents an emerging frontier in magnetic recording
technology \cite{bigotnature10, eric}, where a femtosecond laser pulse
is used to read/write the magnetic information in the storage media
within a few hundred femtoseconds. However, how such an ultrafast
magnetization process occurs is not well understood. 
  The true underlying microscopic mechanism is still
    under debate. Two major mechanims are proposed.  One is the
  Elliot-Yafet (EY) mechanism proposed by Koopmans
  \et\cite{koopmans2005, koopmans2010}, in which the spin relaxation
  occurs via electron scattering at impurities or defects and
  phonons. Further, Steiauf \et\cite{fahnle} invoked a strong
  spin-mixing parameter to support the EY mechanism.  However, the EY
  mechanism is challenged in rare earth compounds as Vahaplar \et
  \cite{Vahaplar} demonstrated a nonthermal writing on a time scale of
  10 ps. On such a long time scale, the phonon should play an
  important role by smearing out any initial polarization dependence
  on the laser light. In a group of the lanthanide-doped permalloys,
  Radu \et \cite{radu} particularly reported an opposite behavior to
  the EY's prediction \cite{koopmans2005}.  Nonthermal nature of
  femtomagnetism was clearly demonstrated in GdFeCo \cite{hohlfeld},
  which also invalidated stimulated Raman scattering
  mechanism. Another mechanism attributes the
  demagnetization to the coherent interaction of the laser beam with
  the electron of the system proposed by Bigot \et
  \cite{bigotnaturephysics09}. In systems with spin-orbit coupling in
  general most of the optical electric dipole transitions of the
  stimulated electrons conserve the dominant spin character of the
  electrons, but a few transitions are spin-flip transitions. The
  number of spin-flip transitions increases with increasing spin-orbit
  coupling. The coherent interaction mechanism is strongly supported
  by the theoretical findings independently developed by Zhang \et
  \cite{prl00, naturephysics09}, and they found that a crucial
  cooperation between laser field and spin-orbit coupling is
  indispensable to ultrafast demagnetization.  In 2009, Lefkidis \et
  \cite{prl09} proposed a spin-switch mechanism in the coherent
  temporal regime, which is based on the angular-momentum exchange
  between the light and the irradiated antiferromagnets.
   Motivated by
  those latest investigations, here we provide a new view on the
  coherent interaction mechanism in femtosecond magnetism.

%    On a much longer time scale, the thermal effect plays a dominant
%    role.  
%Superdiffusive spin transport was also proposed as a mechanism of
%ultrafast demagnetization \cite{oppeneer10}, but based on the results
%of heavy lanthanides, Wietstruk \et \cite{boven10} claimed that their
%observations are not compatible with this mechanism.

  To this end, the existing experimental results are not
  conclusive enough to develop a simple method to characterize the
  spin moment change in femtomagnetism.  In particular, little
  attention has been paid to two most important contributions
  simultaneously: {\it Optical dipole interaction} (ODI) and {\it spin
    expectation value change} (SEC).  SEC is defined as the difference
  between the spin expectation values of two transition states.  ODI
  allows the laser pulse to influence the system, while SEC allows the
  spin moment change during laser excitation.  However, those two
  contributions are intrinsically disconnected. For instance, the
  dipole interaction does not guarantee spin moment change. In optical
  dipole transitions, without spin-orbit coupling, the spin is
  conserved.  The separation of ODI and SEC is unique 
  to femtomagnetism, and is very
  different from the magnetization process driven by a thermal or
  magnetic field, where the spin moment change is the only quantity
  that should be considered.  Therefore, to fully understand
  femtomagnetism, it is a must to take both contributions into account
  simultaneously.

%These experimental and theoretical controversies are not surprising at
%all, since they merely suggest much more involved ultrafast
%demagnetization mechanisms in different materials \cite{boven11}.

%This requires a detailed examination of their electronic and magnetic
%structures.  In this regard, the momentum-resolved scheme is proved
%very powerful. As shown by our group \cite{prb09}, one can see clearly
%where the spin excitation occurs along each high symmetry line (see
%the top panel of Fig. \ref{fig1}), and how the transition matrix
%elements and spin matrix elements jointly determine the amount of spin
%moment change.  The momentum-resolved occupation change, though not
%magnetic moment change, was calculated in two Heusler alloyse
%\cite{steil} lately. One recent experiment \cite{schmidt} again
%demonstrated the value of this spin-, time-, energy- and
%angle-resolved two-photon photoemission but their emphasis is on
%ultrafast magnon generation, not on magnetization change.

\newcommand{\K}{{\bf k}}

 In this Letter we take into account both contributions by introducing
 a new concept - optical spin generator or \OSGe. This generator is
 defined as a product of optical dipole transition moment and SEC for
 a particular transition. If multiple transitions are involved, a
 summation over those transitions should be carried out. We test this
 concept in ferromagnetic nickel. Our first-principles calculation
 shows that \OSG has the capability to single out optically hot spin
 spots. In nickel, two major hot spots are
   identified.  Their structures in the Brillouin zone are almost
   identical to those constructed from the optical spin generator.
   Therefore, our finding not only greatly simplifies the
   interpretation of femtomagnetism, but also points out a new
   direction for future experimental investigations.

We start with a transition from state $|a\ra$ to state $|b\ra$.  The
 expectation value of
optical spin generator {\OSGe} is defined as 
\be \left ( \osg\right)_{ab} \equiv
D_{a\rightarrow b} \Delta S_{a\rightarrow b} +c.c.\ee 
where $D_{a\rightarrow b}$ is a dipole transition moment, $\Delta
S_{a\rightarrow b}=S_b-S_a $ is SEC and $S_a$ refers to the spin
expectation value in state $|a\ra$, and $c.c.$ refers to the complex
conjugation. The direction indices of \OSG are omitted for brevity.
To see an analytic example, we resort to two states characterized by
total angular momentum quantum number $j$ and its magnetic quantum
number $m_j$.   First, we
form the eigenstates for each $j$ and $m_j$. Then, we compute the dipole
transition matrix element and the spin matrix element among those
states. Finally, we compute the product of these two matrix elements.
A lengthy but straightforward calculation \cite{zhangprb09}
shows that for a transition with $\Delta m_j=0$ and $\Delta j=1$, the
expectation value of 
$(\osg)_{ab}$ generator for this transition is 
\be
(\osg)_{ab}=-\frac{1}{2} \sqrt{ \left(1-\frac{m_j^2}{(j+1)^2}\right )
  \left (\frac{m_j}{j(j+1)}\right )^2 }, \ee
 where the radial contribution and its unit are not included.  One
 sees explicitly that 
the expectation value of optical spin generator nonlinearly 
depends on $j$ and $m_j$; such
a nonlinear dependence results from their distinct dependence of spin
and dipole expectation values on $j$ and $m_j$ though 
both spin and dipole changes monotonically with $j$ and $m_j$, a finding 
which is verified in an entirely different system \cite{epl11}.

\begin{figure}
%\begin{minipage}{0.5\textwidth}

\includegraphics[width=7cm]{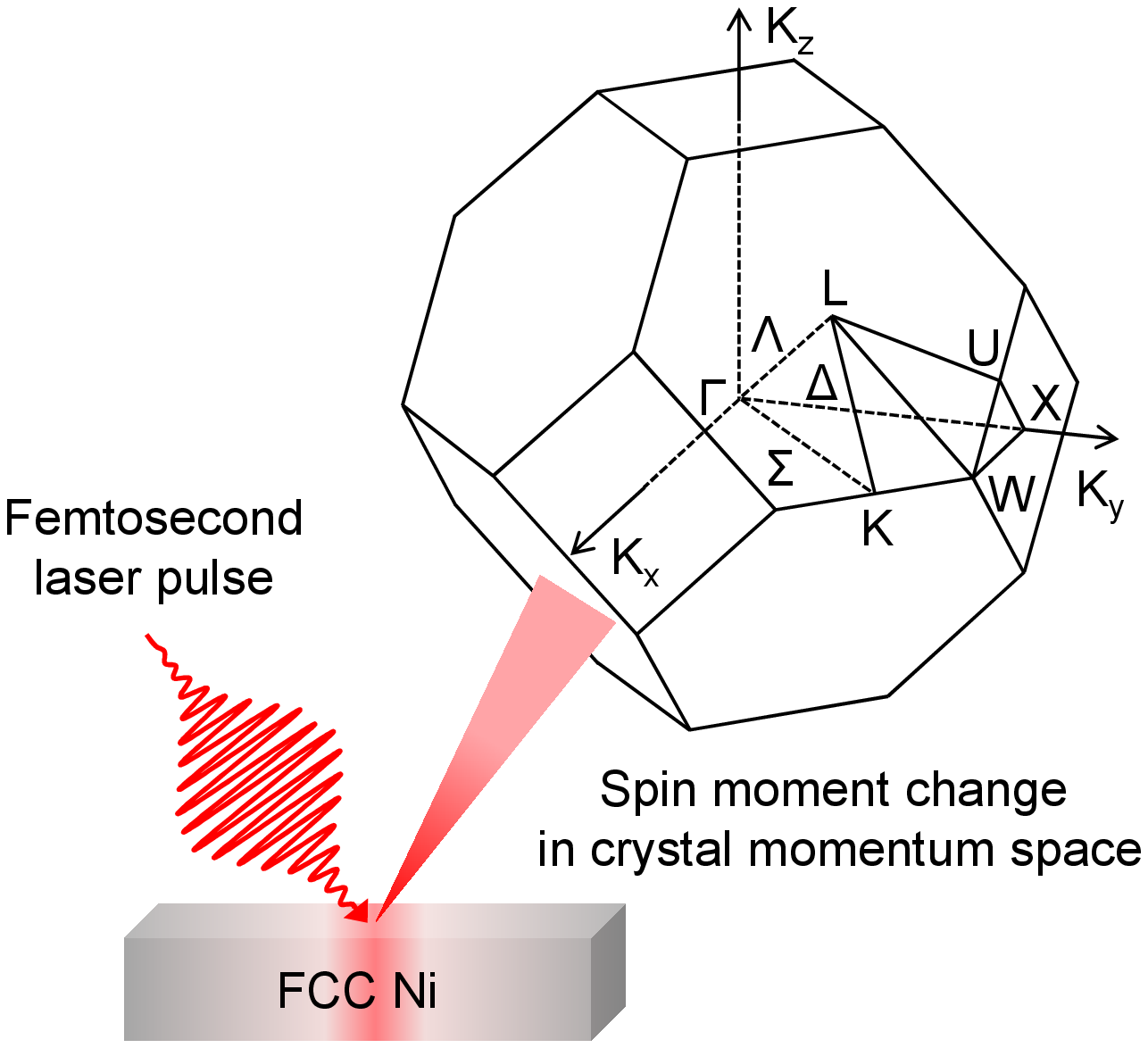}

\includegraphics[width=13cm]{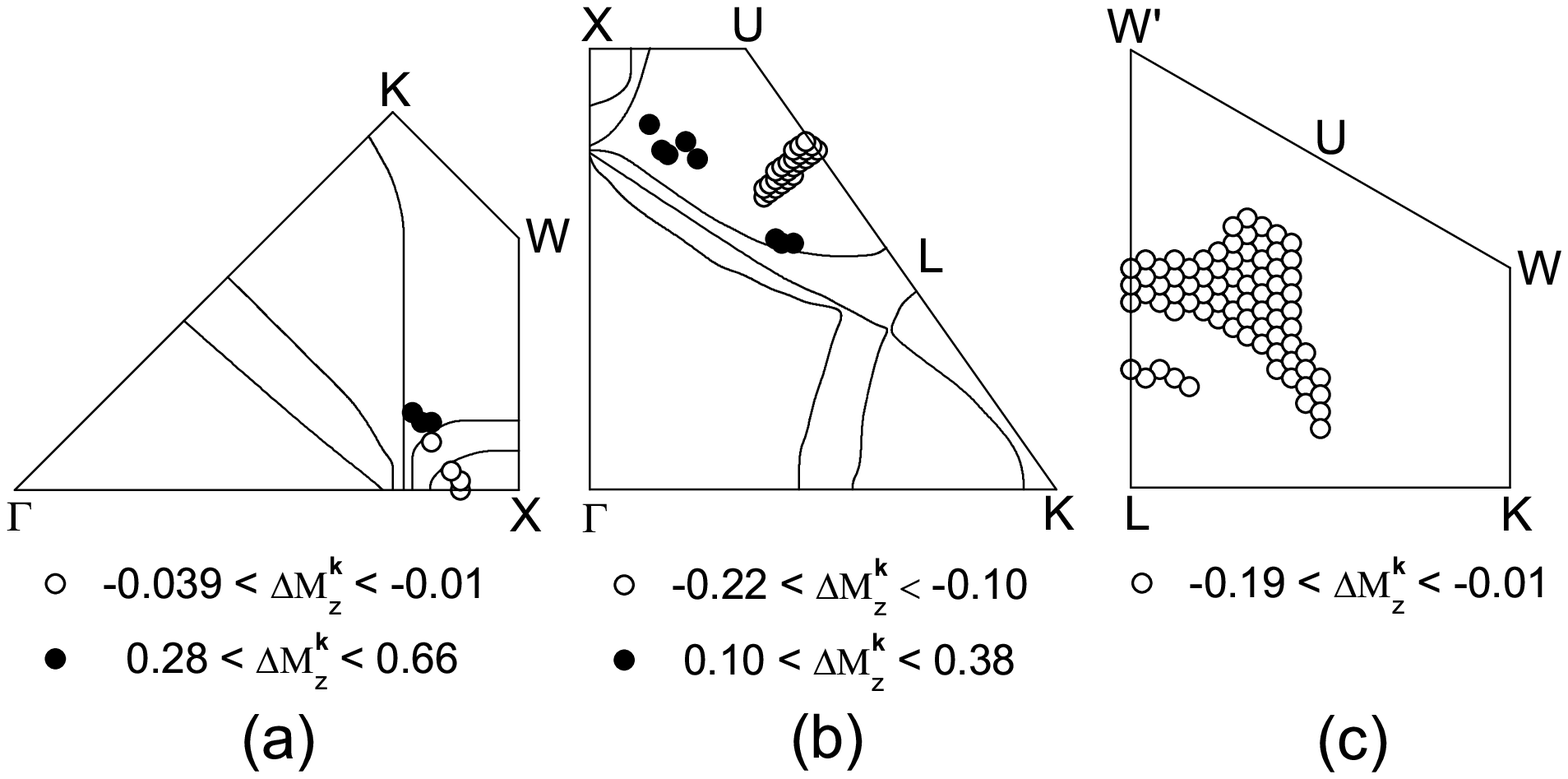}

%\end{minipage}\hfill

\caption{Top: Femtosecond laser induced spin moment change {$\Delta
    M^{\bf k}_z$} is dispersed in the full Brillouin zone in fcc
  Ni. Bottom: (a) Spin moment increase (filled circles) and decrease
  (empty circles) in the $\Gamma$-X-W-K plane.  The range of spin
  moment change is below each figure and in the units of $\ub$.
  The  curves denotes the Fermi surface.  (b) Spin moment change in the
  $\Gamma$-K-L-U-X plane. There are more $k$ points with spin moment
  reduction than those with moment increase.  (c) Spin moment
  reduction dominates the hexagonal L-K-W-U-W$'$ plane.  The Fermi
  surface does not cross through this plane.  }
\label{fig1}
\end{figure}

In the following, we are going to demonstrate the power of optical
spin generator \OSG in ferromagnetic nickel.  While the details of our
theoretical formalism have been presented before
\cite{naturephysics09}, here in brief, we start our calculation
by solving the Kohn-Sham equation self-consistently.  The spin-orbit
coupling (SOC) is included explicitly, without using a spin mixing
parameter \cite{krauss}. Once we obtain eigenstates $(\psi_{n{\bf k}},
{\cal E}_{n{\bf k}})$, we then compute the spin and optical
properties. We find that to converge our spin change, the number of
required \K-points exceeds $87^3$ ($104^3$ used in this study) in the
Brillouin zone, which immediately imposes a big challenge for all the
latter calculations.  To test the accuracy of our results, Figures
\ref{fig1}(a) and (b) show the Fermi surfaces on two high symmetry
planes, where the agreement with the previous calculations
\cite{wang,bunemann} is excellent.  Note that there are two 
subbands - one for each dominant spin orientation
- though they are mixed up due to SOC.

 The interaction between the fs laser pulse field and the system is
 described by $ H_{I}=-{\bf D}\cdot{\bf E}(t).$ Here ${\bf D}=e{\bf
   r}$ is the electric-dipole operator with $e$ being the electron
 charge and ${\bf r}$ the position operator. The dipole matrix
 elements are directly computed from the first-principles method,
 without using the constant matrix element approximation as others
 \cite{krauss}, which has a serious consequence (see below).  The
 laser field ${\bf E}(t)$ peaks at 0 fs and has a Gaussian shape with
 duration of 12 fs and photon energy of 2 eV.  We numerically solve
 the Liouville equation for the electron density matrices $ \rho_{\bf
   k}$ at each \K ~point \cite{jap09}, \be i\hbar\frac{\partial
   \rho_{\bf k}}{\partial t}=[H,\rho_{\bf k}],\ee where
 $H=H_{0}+H_{I}$.  Due to the huge number of \K ~points, all the
 calculations are run in parallel \cite{jap09}.  The spin moment
 change at each \K~point is computed from $\Delta M_{z}^{\bf
   k}(t)=M_{z}^{\bf k}(t)-M_{z}^{\bf k}(-\infty)$, where the spin
 moment at \K~point is $M_{z}^\K (t)={\rm Tr}[\rho_{\bf k}S_{\bf
     k}^{z}]$, and $S_{\bf k}^{z}$ is the $z$-component of spin
 matrix.  Since the spin moment change oscillates with time, we
 time-average it from 90 to 150 fs; averaging the changes from 120 to
 150 fs yields a similar result.

We first show that those conventional high symmetry lines and planes
in the Brillouin zone are not a good guide for spin moment
change. 
Figures \ref{fig1}(a)-(c) show the
crystal-momentum-resolved major magnetic spin moment changes $\Delta
M_{z}^{\bf k}(t)$ on those traditional high symmetry lines and planes,
where the filled circles denote the moment increase and the empty
circles the moment reduction, with quantitative changes shown below
each figure.  In the $\rm \Gamma$-X-W-K plane, the major spin
reductions occur close to the Fermi surface (see Fig. \ref{fig1}(a)),
but in the $\rm \Gamma$-K-L-U-X plane, the major change is away from
the Fermi surface. The maximum reduction reaches
  $-0.22\mu_{\rm B}$. The hexagonal L-K-W-U-W$'$ plane, which does
not intersect with the Fermi surface, has a major change around the
L-U line.  Note that $k$ points with the spin moment increase also
appear in these planes.  The net spin change comes from
  the superposition of the two types hot spots. 

\begin{figure}%[!thb]
%\subfigure{\includegraphics[width=6cm]{./Fig2a.eps}}
%\subfigure{\includegraphics[width=8cm]{./Fig2b.eps}}
%\subfigure{\includegraphics[width=12cm]{./Fig2c.eps}}
\includegraphics[width=14cm]{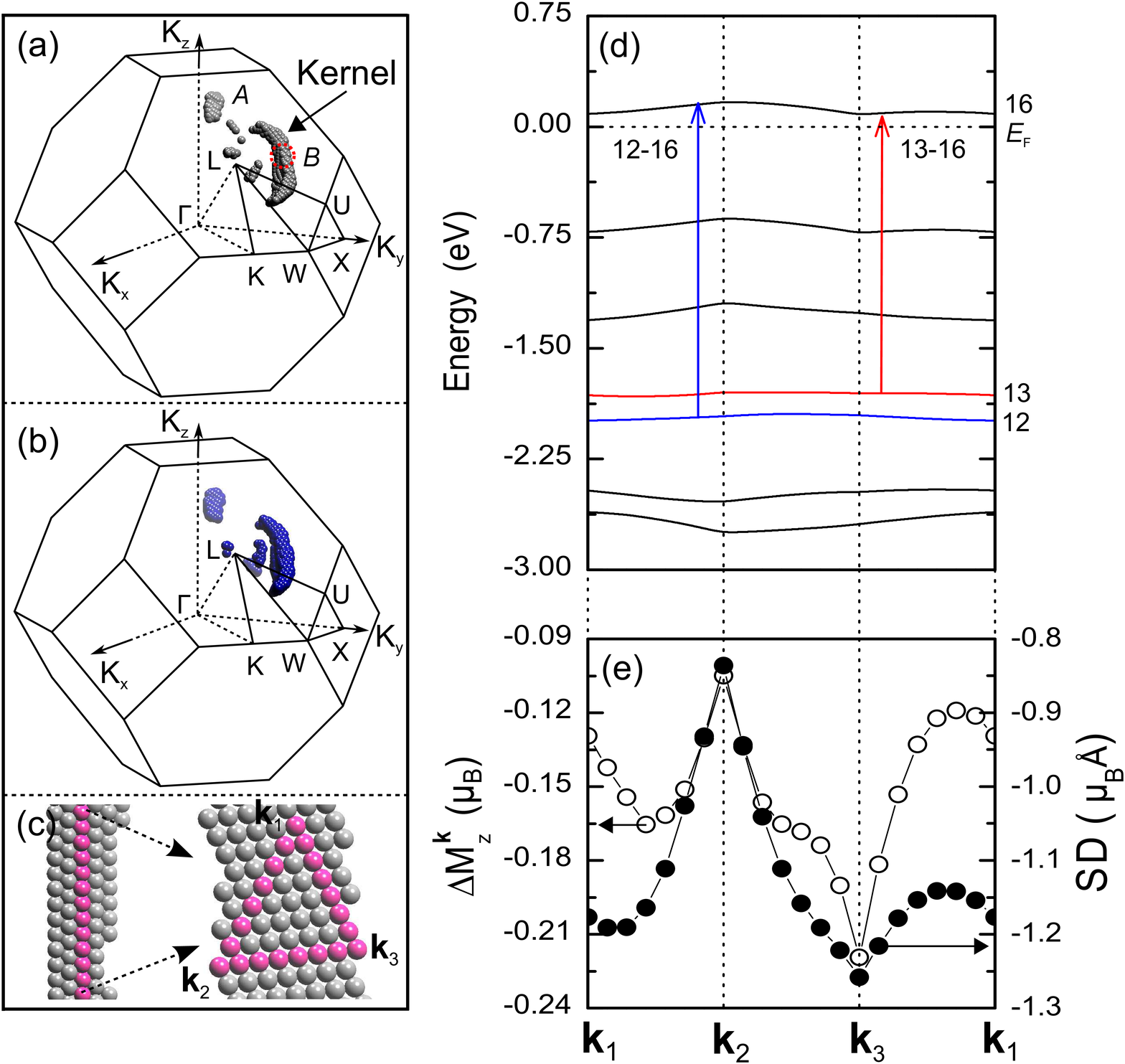}
\caption{\label{fig2}(color online) (a) Hot spin spots
  in the three dimensional Brillouin zone.  Only $k$ points with spin
  moment reduction smaller than $ -0.1 \ub$ (more negative) are shown.
  Two hot spots are labeled by $A$ and $B$. (b) Hot
    spin spots constructed from the maximum optical spin generator
  {\OSGe} reduction smaller than $ -0.13 \ub$\AA.  The number of these
  $k$ points is the same as in (a).  (c) Left: A side-view of spot
  $B$, where a stack of layers are visible.  Right: One layer from
  spot $B$ viewed along the plane normal. The triangle is defined by
  three ${\bf k}_{1}$, ${\bf k}_{2}$, and ${\bf k}_{3}$ points, with
  their respective coordinates of $(27, 73, 49)/104$, $(27, 73,
  35)/104$, and $(34, 80, 42)/104$, in the units of $2\pi$/$a$.  (d)
  Band energy dispersion of the triangle.  The Fermi energy $E_{\rm
    F}$ is set to zero. Two dominant transitions are from level
  $12\rightarrow 16$ (blue arrow) and level $13\rightarrow 16$ (red
  arrow).  (e) Dispersion of the optical spin generator {\OSGe}
  (filled circles) and the spin moment change (empty circles) with
  crystal momentum for the triangle. The maximum reduction $k$ points
  are along the ${\bf k}_1$-${\bf k}_3$ edge.}
  % The maximum reduction of
  %$-0.22 \ub$ is along the ${\bf k}_1$-${\bf k}_3$ edge.  
\end{figure}

Next we demonstrate the power of optical spin generator to
characterize the spin moment change in two steps: (1) we thoroughly
examine the spin moment change $\Delta M_{z}^{\bf k}(t)$ at each ${\bf
  k}$ in the entire Brillouin zone and (2) directly compute the \OSG
and compare it with the real spin moment change.  Figure \ref{fig2}(a)
illustrates the first comprehensive dispersion of $\Delta M_{z}^{\bf
  k}(t)$, which is constructed out of 73763 irreducible $k$ points.
To reduce the huge volume of the data, we only plot 750 $k$ points
with spin moment change $\Delta M_z^{\bf k}$ smaller than $-0.1
\mu_{\rm B}$ (more negative).  This comprehensive dispersion finally
narrows our attention to only two major spin hot spots and some
small islands.  Our findings are insightful.  First, structurally the
hot spin spots consist of stacks of layers of spin hot
points in the crystal momentum space. These hot spots
  with large spin moment reduction are big enough to be detected
experimentally. A small portion of spot B is displayed at the bottom
left of the figure, with one representative layer magnified on the
bottom right. Then we choose a triangle, highlighted by dark red
balls, for a close examination. Choosing such a triangle is purely
arbitrary, and is mainly in the consideration of the continuation of
the coordinates of the \K-points; the conclusion is same when we
choose a different triangle in different layers.  To define the above
triangle, we choose three \K-points, ${\bf k}_1$, ${\bf k}_2$, and
${\bf k}_3$, whose coordinates are given in the caption of
Fig. \ref{fig2}.

Second, different from the single-channel excitation process as seen
in our above analytic example, the laser photon pre-selects multiple
channels.  To see this clearly, we present the energy band dispersion
along three sides of the triangle (see Figure \ref{fig2}(d)).  There
are two transitions with their transition energies close to the laser
photon energy, see two vertical arrows.  According to our previous
investigation \cite{jpcm10}, this leads to a possible resonant
excitation. Therefore, the generator is constructed by summing over
all the major transitions as $ (\osg)_{\bf k}\equiv\sum_{ij} \Delta
S^z_{{\bf k};ij} D_{{\bf k};ij} $, with $i$ and $j$ being the initial
and final state band indices. In our present case, there are two such
transitions.  The results are illustrated in Fig. \ref{fig2}(b). The
 hot spin spots are nicely reproduced by {\OSGe} except
the small islands.  This explicitly shows that the {\OSGe} is able to
characterize the spin moment change of femtomagnetism in
the crystal-momentum space. Figure \ref{fig2}(e) compares the
dispersion of spin moment change with that of {\OSGe}.  We see that
from \K$_1$ to \K$_2$, \OSG drops initially (more negative) and then
increases sharply (less negative) when close to \K$_2$. These changes
are a joint effect of the SEC and dipole moment. As seen in the above
analytic example, the change is highly nonlinear. This is partly due
to the fact that $\Delta S$ and $D$ generally depend on quantum
numbers and ${\bf k}$ points differently. Going from
\K$_2$ to \K$_3$, we see a monotonous decrease (more negative). On the
\K$_3$-\K$_1$ side, \OSG shows a peak. These changes will manifest
themselves in the spin moment change at each ${\bf k}$ point.

\newcommand{\dmz}{$\Delta M^{\bf k}_z$}

Figure \ref{fig2}(e) shows the spin moment change $\Delta M_z^{\bf k}$
along three sides of the same triangle.  It is astonishing that
$\Delta M_z^{\bf k}$ closely follows \OSGe, with all main features
reproduced.  The difference is that \OSG is computed solely from the
SEC and dipole matrix elements while $\Delta M^{\bf k}_z$ is computed
dynamically. Our additional extensive calculations (not included in
the paper) yield the same conclusion.  This proves unambiguously that
\OSG largely governs the nature of laser-induced spin dynamics. If we
compare those two figures more closely, we do see some differences
between $\Delta M^{\bf k}_z$ and \OSGe. For instance, along the
\K$_1$-\K$_2$ side, \OSG has a smaller decrease than that in \dmz;
along the \K$_2$-\K$_3$ side there is a kink in \dmz. Along the
\K$_3$-\K$_1$ side, the spin moment change is also much more
pronounced than that in \OSGe.  These small differences are directly
associated with the laser pulse shape and the photon energy.
Notwithstanding these small difference, the optical spin generator
provides an easy way to evaluate the spin moment change, without
complicated real time simulations.

Conceptually, the optical spin generator is very useful for future
experimental and theoretical investigations, not only in ferromagnets
but also in magnetic semiconductors, since \OSG is essentially a
system quantity, and the only external input to construct \OSG is the
energy window set by laser photon energy.  The physics meaning of \OSG
is very clear: It acts as a source term for the laser to influence
spin.  We emphasize that had we used the constant dipole matrix, \OSG
would be a simple spin matrix, and the optical dipole selection rule
would be ignored. The hot spin spots are a
manifestation of \OSGe. Future experimental research should be able to
test our results.

In conclusion, we have introduced a new concept of the optical spin
generator to simultaneously take into account both the dipole
transition and spin expectation value change between transition
states, each of which is indispensable to laser-induced femtosecond
magnetism. The generator directly links to the spin moment change at
each ${\bf k}$ point. This helps to unfold the hidden kernel of
femtomagnetism in crystal-momentum space. In ferromagnetic nickel, we
find that the hot spin spots consist of two major spin
hot spots and a few smaller islands. These hot spots are not along
any high symmetry lines or plane nor Fermi surfaces. Instead they are
determined by \OSGe. Within these hot spots, the ${\bf
  k}$-dispersed spin moment changes closely follow those of \OSG and
the effect of the laser pulse is reflected from the spectral
transition window pre-selected by the photon energy. This provides a
simple method to characterize the magnetization change on femtosecond
time scale. Future experiments can directly test our prediction by
focusing spin changes at those crystal momenta.

This work was supported by the U. S. Department of Energy under
Contract No. DE-FG02-06ER46304. This work was also supported by the
National Science Foundation of China (NSFC) under No. 10804038, and
the Fundamental Research Fund for the Central Universities and 
Physics and Mathematics of Lanzhou
University.  We acknowledge part of the work as done on Indiana State
University's high-performance computers. This research used resources
of the National Energy Research Scientific Computing Center at
Lawrence Berkeley National Laboratory, which is supported by the
Office of Science of the U.S. Department of Energy under Contract
No. DE-AC02-05CH11231. Our calculation also used resources of the
Argonne Leadership Computing Facility at Argonne National Laboratory,
which is supported by the Office of Science of the U. S. Department of
Energy under Contract No. DE-AC02-06CH11357.

$^{*}$Communication should be directed to GPZ. Email: gpzhang@indstate.edu

\clearpage

\begin{thebibliography}{99}


\bibitem{bigotnature10} C. Boeglin, E. Beaurepaire, V. Halte,
V. Lopez-Flores, C. Stamm, N. Pontius, H. A. D\"urr, and J.-Y. Bigot, 
Nature {\bf 465}, 458 (2010); 
   A.  Kirilyuk, A.  V. Kimel, and Th. Rasing, 
   Rev. Mod. Phys. {\bf 82}, 2731 (2010).

%Ultrafast optical manipulation of magnetic order 

\bibitem{eric}E. Beaurepaire, J.-C. Merle, A. Daunois, and
  J.-Y. Bigot,
 \prl {\bf 76}, 4250 (1996).

\bibitem{koopmans2005}B. Koopmans, J. J. M. Ruigrok, F. Dalla Longa, and W. J. M. de Jonge
 \prl {\bf 95}, 267207 (2005).

\bibitem{koopmans2010} B. Koopmans, G. Malinowski, F. Dalla Longa,
  D. Steiauf, M. F\"ahnle, T. Roth, M. Cinchetti, M. Aeschlimann,
   Nature Mater. {\bf 9}, 259
  (2010); M. G. M\"unzenberg, {\bf 9}, 184 (2010).
%Distinguishing the ultrafast dynamics of spin and orbital moments in solids

\bibitem{fahnle}D. Steiauf and M. F\"ahnle,
Phys. Rev. B {\bf 79},
140401(R) (2009).


\bibitem{Vahaplar}K. Vahaplar, A. M. Kalashnikova, A. V. Kimel,
 D. Hinzke, U. Nowak, R. Chantrell, A. Tsukamoto, A. Itoh,
A. Kirilyuk, and Th. Rasing,
  Phys. Rev. Lett. {\bf 103}, 117201 (2009).

\bibitem{radu}I. Radu, G. Woltersdorf, M. Kiessling, A. Melnikov,
  U. Bovensiepen, J.-U. Thiele, and C. H. Back,
  \prl {\bf 102}, 117201 (2009).

\bibitem{hohlfeld}J. Hohlfeld, C. D. Stanciu, and A. Rebei,
  Appl. Phys. Lett. {\bf 94}, 152504 (2009).

\bibitem{bigotnaturephysics09}J.-Y. Bigot, M. Vomir, and
  E. Beaurepaire,
 \np {\bf 5}, 515 (2009).

\bibitem{prl00}G. P. Zhang and W. H\"ubner, 
  \prl {\bf 85}, 3025 (2000).

\bibitem{naturephysics09}G. P. Zhang, W. H\"ubner, G. Lefkidis,
  Y. H. Bai, and T. F. George,
  \np {\bf 5}, 499 (2009).

\bibitem{prl09}G. Lefkidis, G. P. Zhang, and W. H\"ubner, 
\prl {\bf  103}, 217401 (2009).  


%\bibitem{oppeneer10} M. Battiato, K. Carva, and P. M. Oppeneer,
%  Phys. Rev. Lett. {\bf 105}, 027203 (2010).
%Superdiffusive Spin Transport as a Mechanism of Ultrafast Demagnetization



%\bibitem{boven10}M. Wietstruk \ete, \prl {\bf 106}, 127401 (2011).

%Strong photo-induced ehancement of spin-lattice coupling in 4f
%ferromagnets observed by femtosecond x-ray magnetic circular dichroism.
%
%


%\bibitem{boven11}M. Sultan, A. Melnikov and W. Bovensiepen,
%  xxx.lanl.gov: arXiv1103.1512v1 (2011).

%Ultrafast magnetization dynamics of Gd (0001): Bulk vs. surface. 
%



\bibitem{zhangprb09}G. P. Zhang, Y. H. Bai, and T. F. George, 
\prb {\bf  80}, 214415 (2009).

\bibitem{epl11}G. P. Zhang, M. S. Si and T. F. George,
  Europhys. Lett. {\bf 94}, 17005 (2011). 


%\bibitem{steil} D.  Steil \ete, \prl {\bf 105}, 217202 (2010).

%Band-Structure-Dependent Demagnetization in the Heusler Alloy Co2Mn1-xFexSi

%\bibitem{schmidt}A. B. Schmidt \ete, \prl {\bf 105}, 197401 (2010).

% M. Pickel, M. Donath, P. Buczek,
%  A. Ernst, V. P. Zhukov, P. M. Echenique, L. M. Sandratskii,
%  E. V. Chulkov, and M. Weinelt, 
%
%Ultrafast Magnon Generation in an Fe Film on Cu(100)



%\bibitem{stamm}C. Stamm, N. Pontius, T. Kachel, M. Wietstruk,
%  H. A. D\"urr, Phys. Rev. B {\bf 81}, 104425 (2010). 

%Femtosecond x-ray absorption spectroscopy of spin and orbital angular
%momentum in photoexcited Ni films during ultrafast demagnetization




%\bibitem{mansuripur}M. Mansuripur, {\it The Physical Principles of
%  Magneto-Optical Recording} (Cambridge University Press, Cambridge,
%  1995).


%\bibitem{zutic}I. \v{Z}uti\'c, J. Fabian, and S. Das Sarma,
%  Rev. Mod. Phys. {\bf 76}, 323 (2004).



\bibitem{krauss}M. Krau\ss, T. Roth, S. Alebrand, D. Steil,
   M. Cinchetti, M. Aeschlimann, and H. C. Schneider, 
   Phys. Rev. B {\bf  80}, 180407(R) (2009).

%Ultrafast demagnetization of ferromagnetic transition metals: The role of the Coulomb interaction



\bibitem{wang}C. S. Wang and J. Callaway,
 \prb {\bf 9}, 4897 (1974).

\bibitem{bunemann}J. B\"unemann, F. Gebhard, T. Ohm, S. Weiser, and
  W. Weber, 
\prl {\bf 101}, 236404 (2008).

%F. Gebhard, T. Ohm, S. Weiser, and
%  W. Weber, 

%Spin-Orbit Coupling in Ferromagnetic Nickel

\bibitem{jap09} T. Hartenstein, G. Lefkidis, W. H\"ubner, G. P. Zhang,
  and Y. Bai, 
J. Appl. Phys. {\bf 105}, 07D305 (2009); G. P. Zhang,
  Y. Bai, W. H\"ubner, G. Lefkidis, and T. F. George,
  J. Appl. Phys. {\bf 103}, 07B113 (2008).


\bibitem{jpcm10}M. S. Si and G. P. Zhang,
 \jpcm {\bf 22}, 076005 (2010).



%\bibitem{supple} http://www.aip.org/pubservs/epaps.html.



%\bibitem{zhang} Q. Zhang \ete, Phys. Rev. B {\bf 74}, 064414 (2006).


%Electronic structure of CrO2 as deduced from its magneto-optical Kerr
%spectra


%\bibitem{muller}G. M. M\"uller \ete, Nature Mater. {\bf 8}, 56  (2009). 

%Spin polarization in half-metals probed by femtosecond spin excitation







%\bibitem{satoh}T. Satoh \ete,
%Phys. Rev. Lett. {\bf 105}, 077402 (2010).

%Spin Oscillations in Antiferromagnetic NiO Triggered by Circularly
%Polarized Light
%

% Phys. Rev. Lett. 106, 047401 (2011) [4 pages]
%Selection Rules for Light-Induced Magnetization of a Crystal with
%Threefold Symmetry: The Case of Antiferromagnetic NiO
%Abstract
%References
%No Citing Articles
%Download: PDF (1,302 kB) Export: BibTeX or EndNote (RIS)

%Takuya Higuchi1, Natsuki Kanda1, Hiroharu Tamaru2, and Makoto
%Kuwata-Gonokami1,2,3,* 


%\bibitem{stanciu}C. D. Stanciu \ete, \prl {\bf 99}, 047601 (2007).




\end{thebibliography}
\end{document}